# Magnetic order of tetragonal CuO ultra-thin films

**N. Ortiz Hernandez[1], Z. Salman[2], T. Prokscha[2], A. Suter[2], J. R. L. Mardegan[3,1], S. Moser[4,5], A. Zakharova[1], C. Piamonteze[1], und U. Staub[1*]**

*1 Swiss Light Source, Paul Scherrer Institute, 5232 Villigen PSI, Switzerland.*
*2 Laboratory for Muon Spin Spectroscopy, Paul Scherrer Institute, 5232 Villigen PSI, Switzerland.*
*3 Deutsches Elektronen-Synchrotron,* 22607 Hamburg, Germany.
*4 Advanced Light Source, E. O. Lawrence Berkeley National Laboratory, Berkeley, California 94720, United States.*
*5 Physikalisches Institut and Würzburg-Dresden Cluster of Excellence ct.qmat, Universität Würzburg, 97074 Würzburg, Germany.*

*Corresponding author: urs.staub@psi.ch

**Abstract**

We present a detailed low-energy muon spin rotation and x-ray magnetic circular dichroism (XMCD) investigation of the magnetic structure in ultra-thin tetragonal (T)-CuO films. The measured muon-spin polarization decay indicates an antiferromagnetic (AFM) order with a transition temperature higher than 200K. The XMCD signal obtained around the Cu $L_{2,3}$ edges indicates the presence of pinned $Cu^{2+}$ moments that are parallel to the sample surface, and additionally, isotropic paramagnetic moments. The pinning of some of the Cu moments is caused by an AFM ordering consisting of moments that lie most likely in plane of the film. Moreover, pinned moments show a larger orbital magnetic moment contribution with an approximate ratio of $m_{orb}/m_{spin}$ = 2, indicating that these spins are located at sites with reduced symmetry. Some fractions of the pinned moments remain pinned from an AFM background even at 360K, indicating that $T_N$ > 360K. A simple model could explain qualitatively these experimental findings; however, it is in contrast to theoretical predictions, showing that the magnetic properties of ultra-thin T-CuO films differ from bulk expectations and is more complex.

## 1 Introduction

The discovery of high-temperature superconducting cuprates (HTSC) [1]-[2] has increased the attention in materials based on different Cu-O building blocks. It was realized that electron correlations in two-dimensional $CuO_2$ planes are responsible for the unusual properties of these class of materials. Binary CuO was studied as a prototypical insulating parent compound in view of its simpler stoichiometry [3]. Unfortunately, bulk CuO crystallizes in a complex monoclinic crystal structure called tenorite, which does not contain the archetypal $CuO_2$ plaquettes with Cu-O-Cu bond angle of 180° [4]. This low symmetry crystal structure makes it difficult to use the properties of bulk CuO as a proxy for the understanding of the HTSC parent compounds.

Later, a higher-symmetric tetragonal distorted Rock-Salt form of CuO (T-CuO) has been synthesized as ultrathin films of approximately 6 unit cells (u.c.) grown on top of $SrTiO_3$ substrate [5]. In this material, $CuO_6$ octahedra forms two-dimensional CuO planes with both edge- and corner-sharing Cu-O bonds that are stacked along the crystal *c*-axis. At variance with the $CuO_2$ cuprate layers, oxygen ions do not bridge nearest neighbor (NN) but next-NN copper ions [6]-[7]. The electronic structure of T-CuO is determined by the properties of its 2D edge-sharing lattice planes, each conceptually built from two interpenetrating corner-sharing sub-lattices, with lattice parameters typical of HTSCs and 180° Cu-O-Cu bonds. A small coupling between these frustrated sub-lattices gives rise to slight quasiparticle renormalization [8]. In addition, T-CuO extends the simple rock salt transition metal oxides series from FeO, CoO, NiO to CuO and has been predicted to have a very high magnetic ordering temperature, with Néel temperature ($T_N$) close to 800 K [9]-[10]. Therefore, T-CuO is a very useful model system to test theoretical first principles calculations on magnetic interactions. A study using CuO-$Cu_2O$ heterostructure with highly symmetric CuO phase claims to find an antiferromagnetic order below $T_N$ = 600 K [11]. Nevertheless, the study of ref. [11] has the clear shortcoming that it claimed not being able to distinguish between the two structural phases $Cu_2O$ and cubic CuO through an XRD (X-Ray powder Diffraction) analysis. The lattice parameter for $Cu_2O$ is approximately 4.28 $\dot{A}$ [12] and the lattice parameter for the cubic CuO phase can be estimated from the average lattice parameters of the T-CuO synthetized by Siemons [5], which is approximately 4.38 $\dot{A}$. The 0.1 $\dot{A}$ lattice parameters difference would be easily detected and sufficient to distinguish between the two structures in an XRD analysis for angles larger than 50 deg. However, the work published in the literature [11] does not find such a contribution and therefore the presence of a cubic CuO phase has not been demonstrated, which makes the discovery related to $T_N$ ambiguous.

Experimentally, resonant inelastic x-ray scattering (RIXS) measurements reported spin wave excitations dispersing through two cuprate-like AFM sublattices. Two possible scenarios for the magnetic structure have been proposed with AFM orderings along the film plane [8]. However, there is so far no direct experimental confirmation of a magnetic order in T-CuO and its possible spin structure, since magnetic excitations away from the zone center have also been found in superconducting cuprates by RIXS at doping levels far from the antiferromagnetically ordered state [14]. Here, we perform a series of complementary low energy muons spin rotation (LE-µSR) and x-ray magnetic circular dichroism (XMCD) measurements on T-CuO ultra-thin films. Our results present clear evidence of static magnetic order in T-CuO. XMCD finds pined magnetic moments, which point to an underlying AFM as well as some paramagnetic contribution from the interface or surface region. These pinned moments clearly suggest an AFM ordering which preserves above room temperature but are difficult to understand in context with the predicted magnetic ordering schemes of T-CuO.

# 2 Experiments

## 2.1 Sample preparation

Four nominally identical T-CuO thin films have been epitaxially grown on the (001) surface of a $SrTiO_3$ (STO) substrates with dimensions 10x10 mm by pulsed laser deposition, following the same procedure reported in ref. [5]. The T-CuO layer thickness was estimated to 3.6 nm (approximately 6 u.c.), in which it was controlled by reflection high-energy electron diffraction. A thin carbon layer of 10 nm thickness was deposited on top of the T-CuO films at room temperature and at $10^{-5}$ mbar pressure by a commercial LEICA EM ACE600 high vacuum sputter coater, which measures in-situ the thickness of the capping C layer. These samples are referred as C/T-CuO/STO. The carbon layer serves as a protection layer for the T-CuO without affecting the properties of the thin film. Additionally, four (001)-STO bare substrates of 10x10 mm were coated in the same way and used as reference. The coated substrates are labelled as C/STO.

## 2.2 Low energy muons spin rotation (LE-μSR)

The LE-μSR experiments have been performed in the LEM spectrometer [13]-[15] at the Paul Scherrer Institute. One of the great advantages of the LEM spectrometer is that it can implant 100% spin polarized positively charged muons ($\mu^{+}$) in the investigated material. Due to the very thin layer of T-CuO, implantation energies from 1 to 6 keV were selected for this investigation. A static magnetic field of 100 G, perpendicular to the sample surface and initial muon polarization was applied during the measurements. An array of eight positron detectors placed around the sample measures the decay product of the muons, i.e., positrons, as a function of time. The asymmetry in the number of detected positrons between two opposite detectors is proportional to the spin polarization of the muons along their axis [16]. For the LE-μSR experiments, the four C/T-CuO/STO samples were arranged in a mosaic covering an area of 20 x 20 mm and glued using silver paint onto a silver coated aluminum plate. Same procedure was carried out with the four C/STO reference samples. The plate was then mounted onto a cold finger of a He flow cryostat, which operates in the temperature range from 5 to 300 K. Due to the large beam area at the LEM spectrometer, only ~75% of the muons stop in the sample itself while ~25% stop in the silver coating and contribute a non-relaxing background signal.

## 2.3 X-ray magnetic circular dichroism (XMCD)

The XMCD data were collected at the X-TREME beamline [17] of the Swiss Light Source of the Paul Scherrer Institute. An x-ray beam with spot size of 1 x 1.3 mm was used. The setup at X-TREME allows to field cool (FC) the sample from 360 K to 3 K in a magnetic field ranging from 6.5 T (+FC) to -6.5 T (-FC). The magnetic field was always parallel to the incident x-ray beam. Incident angles for the x-rays of 30 and 90 degrees from the sample surface were selected, ensuring enhancements of the in-plane and out-of-plane components of the magnetization in the XMCD signal, respectively. Measurements collected in total electron yield mode were carried out with the photon energies tuned around Cu $L_{2,3}$ absorption edge. One of the C/T-CuO/STO samples used for the LE-μSR experiments was used for the XMCD measurements. Additionally, x-ray linear dichroism (XLD) measurements were also performed at room temperature and with a 30 degrees incident angle.

# 3 Results

## 3.1 Low energy muons spin rotation (LE-μSR)

Low energy muons are highly sensitive probes of the local magnetic fields in the investigated sample. Thanks to their tunable implantation energy (E), and corresponding implantation depth on nanometer scale, the muons can be used to study very thin samples. The implanted muons decay emitting positrons ($e^+$), preferentially, in the muons spin direction at the time of decay. By detecting the time dependence of the positron counts in different detectors ($N_{e^+}(t)$), we can determine the spin polarization of the muon ($P(t)$), as [18]

$$N_{e^+}(t) = N_0 \left(1 + A_\mu P(t)\right) e^{-t/\tau_\mu} \tag{1}$$

where $N_0$ is the total number of recorded positrons at t=0 and $A_\mu$ is the asymmetry parameter, which depends on the anisotropy of the beta decay of the $\mu^+$ and the geometry of the detector. The exponential decrease in $N_{e^+}(t)$ is due to the muon lifetime $\tau_\mu$= 2.2 μs. In the presence of a constant magnetic field, applied transverse to the initial muon spin polarization, $P(t)$ precesses due to the influence of the magnetic field following,

$$A_\mu P(t) = A_\mu e^{\frac{-\sigma^2 t^2}{2}} cos(\omega_\mu t + \varphi) \tag{2}$$

The Larmor frequency, $\omega_\mu = \gamma_\mu B$, where $\gamma_\mu = 0.8516\, Mrad/mT$ is the muon gyromagnetic ratio and $B$ is the local magnetic field detected by muons, which corresponds to the sum of the local magnetic field of the sample and the external applied magnetic field. The Gaussian envelope in Eq. 2 represents the damping of the oscillating polarization with relaxation parameter $\sigma$ due to the width of the local field distribution. The phase $\varphi$ depends on the relative orientation of the initial polarization and the positron detector. In insulators materials, muonium formation occurs when a muon binds an electron forming a hydrogen like state ($\mu^+$+$e^-$).

In order to determine and optimize the fraction of muons implanted in the very thin T-CuO film, Trim.SP Monte Carlo simulations [18]-[19] were performed (Figure 1). Figure 1 shows that with an implantation energy of 1.5 keV, a 10 nm carbon-capping layer on top of T-CuO deaccelerates the muons and results in an optimized stopping fraction of muons in T-CuO of 20-30 %. For this purpose, the four T-CuO thin films have been coated with 10 nm carbon layer (C/T-CuO/STO) and additionally, also the reference C/STO samples.

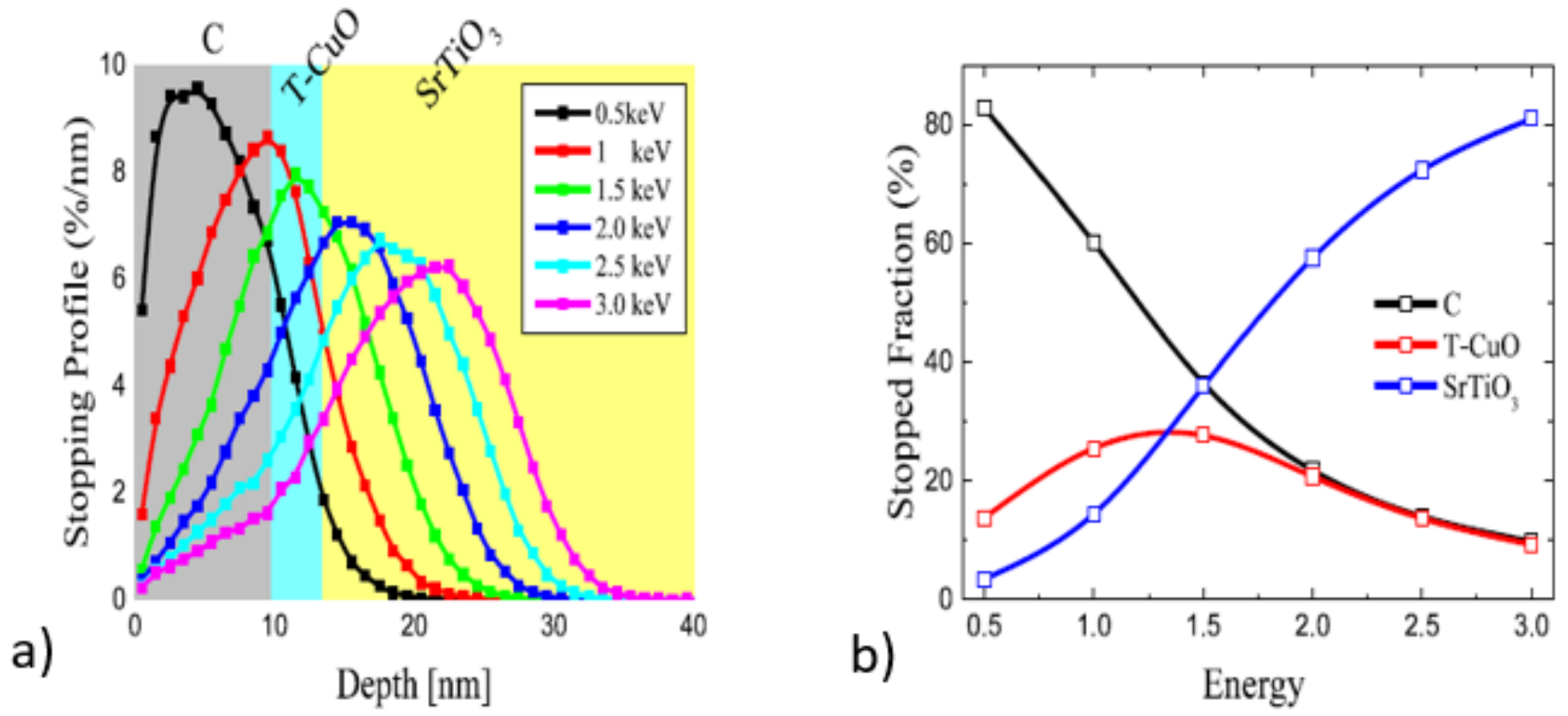

*Figure 1. Results of Trim.SP simulation to calculate the stopping distribution of muons in C/T-CuO/STO. (a) Stopping profile as a function of depth for different implantation energies. (b) Stopped fraction of muons, for each layer, as a function of implantation energy.*

The asymmetry as a function of temperature was measured in both samples at various implantation energies. The data have been fitted using Eq. 2 to evaluate the relevant parameters. Figure 2a displays the temperature dependence of the $A_\mu$ measured in the C/T-CuO/STO sample for different implantation energies. At 5 keV, the temperature dependence clearly resembles to that observed in bare STO [20]. The variation in temperature around 90 K is attributed to the muonium formation in the STO substrate that accounts for the "lost" diamagnetic asymmetry at low temperatures. With decreasing the E, we expect an increase in the contribution from the T-CuO layer. Data with implantation energy of 2 keV shows that the signal from the bare STO still remains. At 1.5 keV a clear deviation from bare STO becomes more apparent as expected from the Trim.SP simulation (Figure 1). This difference is clearly seen when comparing $A_\mu$ measured in C/STO and C/T-CuO/STO as a function of implantation energy, as shown in Figure 2b. Note that these measurements were performed at high temperatures, where there is no muonium formation in STO. At high implantation energy (i.e. when muons stop in the substrate) the asymmetry measured in both samples is equal as expected for samples of similar size and composition. However, decreasing E, a gradually difference between the asymmetries of both samples is observed, where C/T-CuO/STO exhibits a larger reduction. We attribute the reduced asymmetry in C/T-CuO/STO to a fast depolarization of muons implanted in the T-CuO layer. This fast depolarization is a strong indication of a large local magnetic field present in this layer, which is already presented at 200 K.

Figure 2c-e display the parameters $A_\mu$, the mean field $B$ and the damping rate $\sigma$ extracted from the fits using Eq. 2 as a function of temperature for the C/STO and the C/T-CuO/STO samples at E = 1.5 keV. The asymmetry in Figure 2c for the C/T-CuO/STO sample reduces by approximately 6% compared to the C/STO in the full temperature range. This reduction is consistent with the energy dependence displayed in Figure 2b and presents a clear sign of magnetic order in T-CuO layer. The mean field B (Figure 2d) exhibits no temperature dependence and no significant difference between the C/T-CuO/STO and the C/STO samples. In contrast, the damping rate $\sigma$ (Figure 2e) shows a small enhancement in C/T-CuO/STO at low temperatures compared to C/STO.

To understand the nature of magnetic order in the T-CuO layer, we note first that the fraction of muons stopping in the T-CuO predicted by the Trim.SP simulations is ~25% at 1.5 keV, as shown in Figure 1. Considering that only ~75% of the muons contribute to the signal from the C/T-CuO/STO sample (the remaining ~25% stop in the silver coating) and assuming that all the muons implanted in the T-CuO experience a large magnetic field and depolarize, we would expect a loss of asymmetry of ~18% in C/T-CuO/STO compared to C/STO. However, we observe approximately 6% symmetry reduction, which is just 1/3 of the asymmetry reduction expected. This indicates that only 1/3 of the T-CuO volume exhibits long-range magnetic order. Nevertheless, this estimation is very crude since it is based on Trim.SP calculations and takes the bulk densities of C and T-CuO, which might be significantly different in the studied heterostructure.

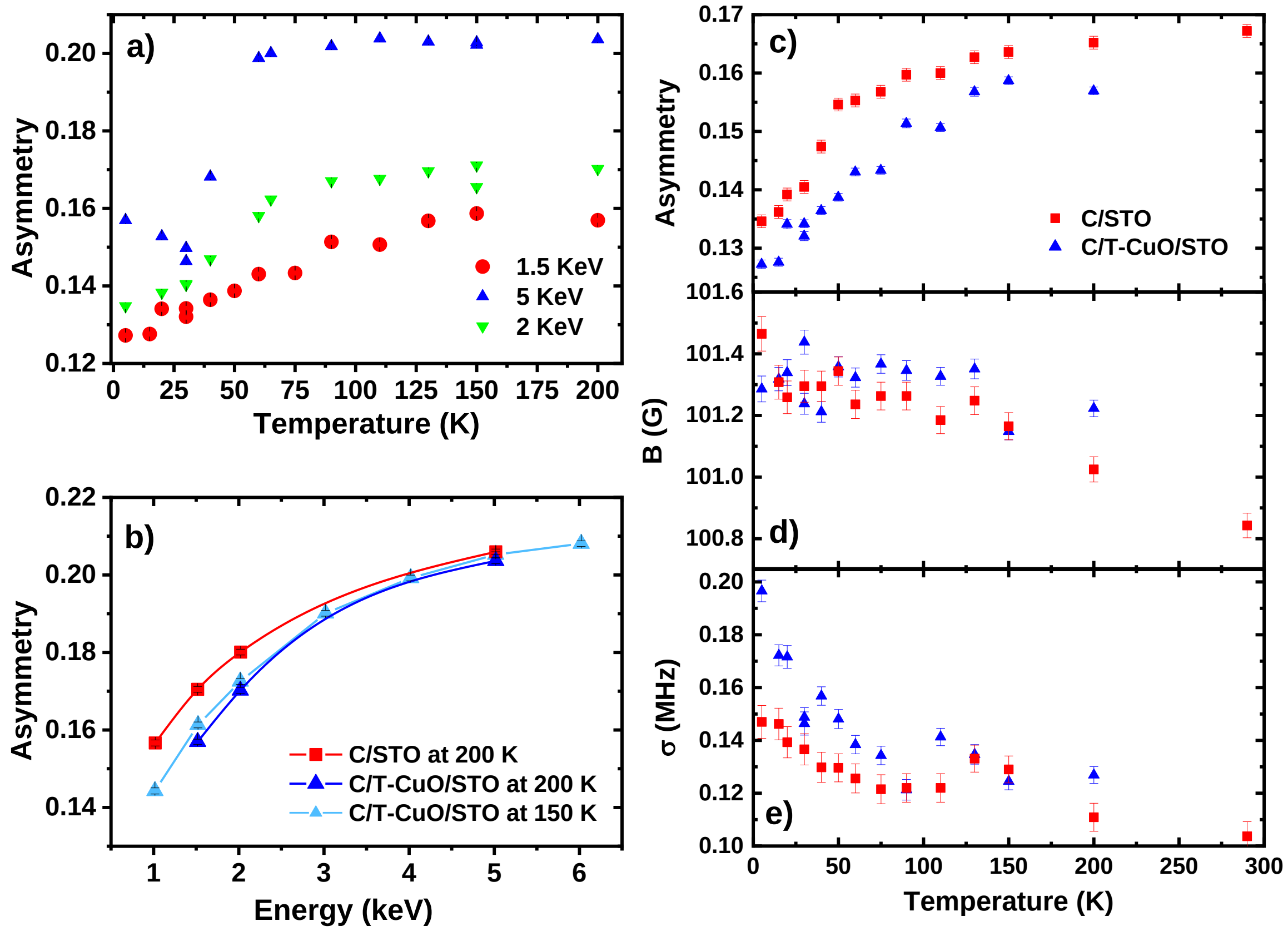

*Figure 2. (a) Temperature dependence of the asymmetry for three different muon implantation energies for C/T-CuO-STO sample. (b) Measured asymmetry as a function of implantation energy of C/STO at 200 K and C/T-CuO/STO at 200 K and 150 K (lines are guides to the eye). (c) Asymmetry, (d) field and (e) damping rate extracted from fits of the data measured as a function of temperature in the two specimens, C/STO and C/T-CuO/STO at*

## 3.2 X-ray magnetic circular dichroism (XMCD)

XMCD is the difference between the absorption spectra taken with right and left circularly polarized light (C+ and C-). The dichroic signal is non-zero for systems with a net magnetization, such as ferromagnets, uncompensated AFM, paramagnets in applied fields, etc. In addition, XMCD is directly proportional to the magnetic moment component that is parallel to the x-ray beam direction. It should be noticed that a perfect compensated AFM system will therefore not create an XMCD signal, since the magnetic moment is fully compensated. However, we found a clear XMCD signal as our following results show.

Figure 3 shows the X-ray Absorption Spectrum (XAS) and the XMCD signal for +FC and applied field of 6.5 T in panel (a-b) and -6.5 T in panel (c-d), respectively, both taken at 3 K and covering the energy range of the Cu $L_{2,3}$ edges. The XAS has been represented as $(\mu^+ + \mu^-)/2$ and the electron transitions to the continuum has been extracted by a two-step function. The XMCD signal is $(\mu^+ - \mu^-)$. The measurements were performed with an incident angle of 30 deg. We detect a much larger XMCD signal for the (b) case with 6.5 T applied compared to (d) case with -6.5 T applied. This indicates that a fraction of magnetic $Cu^{2+}$ moments pointing parallel to +H direction do not reverse their orientation in the opposite magnetic field at this temperature. This can be attributed to pinned moments [21]. In addition, we observe a relative increase of the $L_2/L_3$ XMCD intensity ratio for the (d) case.

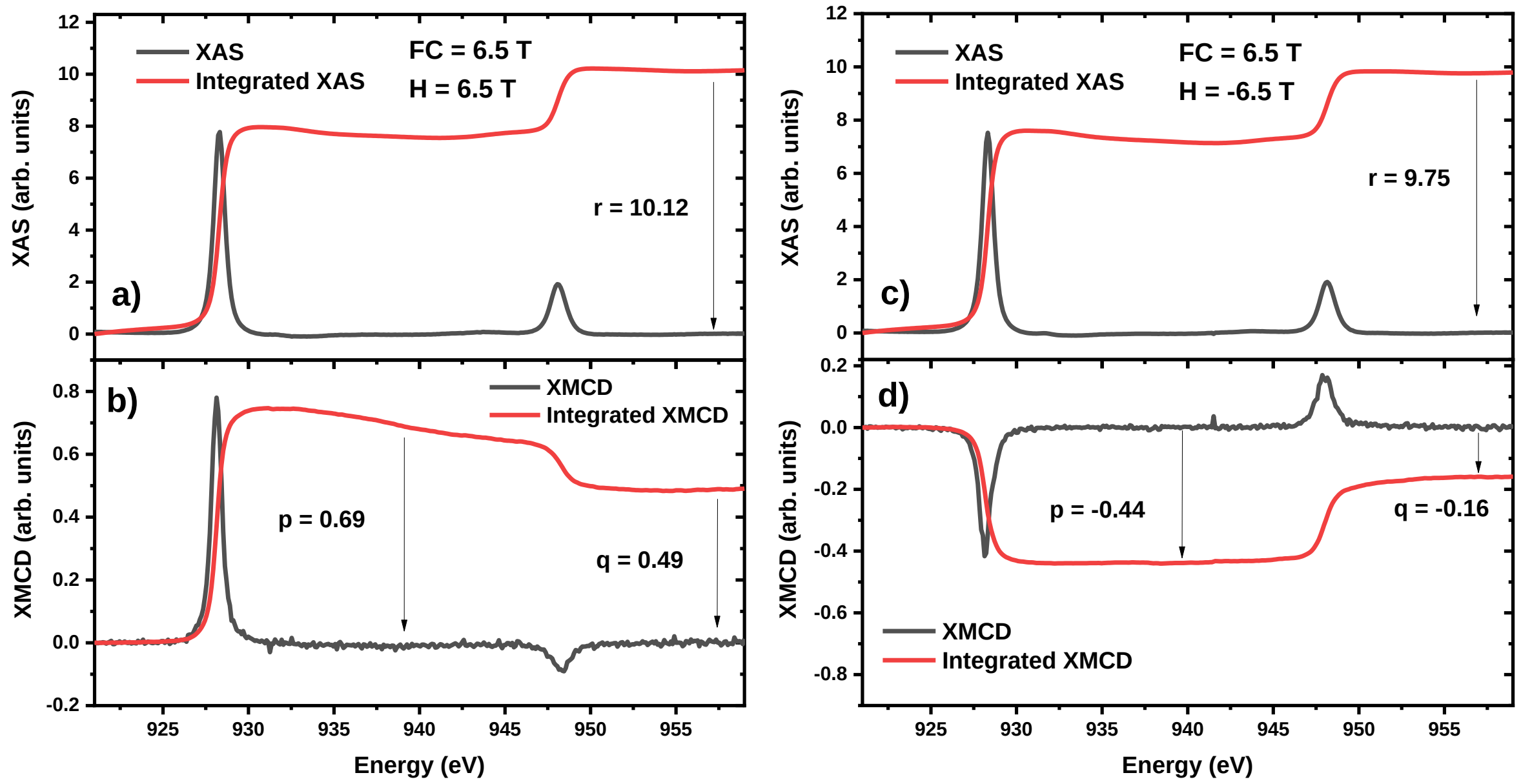

*Figure 3. XAS and XMCD signals at 3K for positive FC (6.5 T) and applied magnetic field of (a -b) 6.5 T and (b-c) - 6.5 T, respectively. Data collected at 30 degrees incident angle.*

One can estimate, through XMCD sum rules, the orbital ($m_{orb}$) and spin moment ($m_{spin}$) per Cu ion of our system as [22]-[23]:

$$m_{spin} = \frac{-3p - 2q}{r}\left(1 + \frac{7 < T_z >}{2 < S_z >}\right)^{-1} \tag{3}$$

$$m_{orb} = \frac{-2q}{3r} N_h \tag{4}$$

Where $p$ is the integral of the XMCD signal over the Cu $L_3$ edge, $q$ is the integral of the XMCD signal over the $L_3$ and $L_2$ edges, $r$ the integral of the isotropic spectra over $L_{2,3}$ edges as indicated in Figure 3. $T_z$ is the dipolar term and $N_h$=1 is the number of holes in the 3$d$ shell for the $Cu^{2+}$ state. In cases where the ion is in a high-symmetry site like an octahedral site, $T_z$ can be neglected, but for T-CuO with a tetragonal symmetry, it might not be negligible [24]. We calculated $T_z$ using the Crispy software [25] that is based on the Quanty code [27]. The crystal field parameters were taken from Moser et al. [8] with 10Dq = 1.5 eV, Ds= 0.25 eV and Dt= 0.13 eV. The calculated X-ray Linear Dichroism (XLD) spectra is in a good agreement with our experimental XLD data (Figure 4). The XLD is the difference between horizontal and vertical linear polarization, and it is very sensitive to the crystal field splitting. From the calculation, we obtain a value of $T_z$ = 0.016, $S_z$=-0.494 and $L_z$=-0.121. The total magnetic moment ($S_z$+$L_z$) obtained by multiplet calculations is of comparable size to those reported by cupric oxides systems [26]. The XMCD sum rules results using eq. 3 and 4 are listed in Table I.

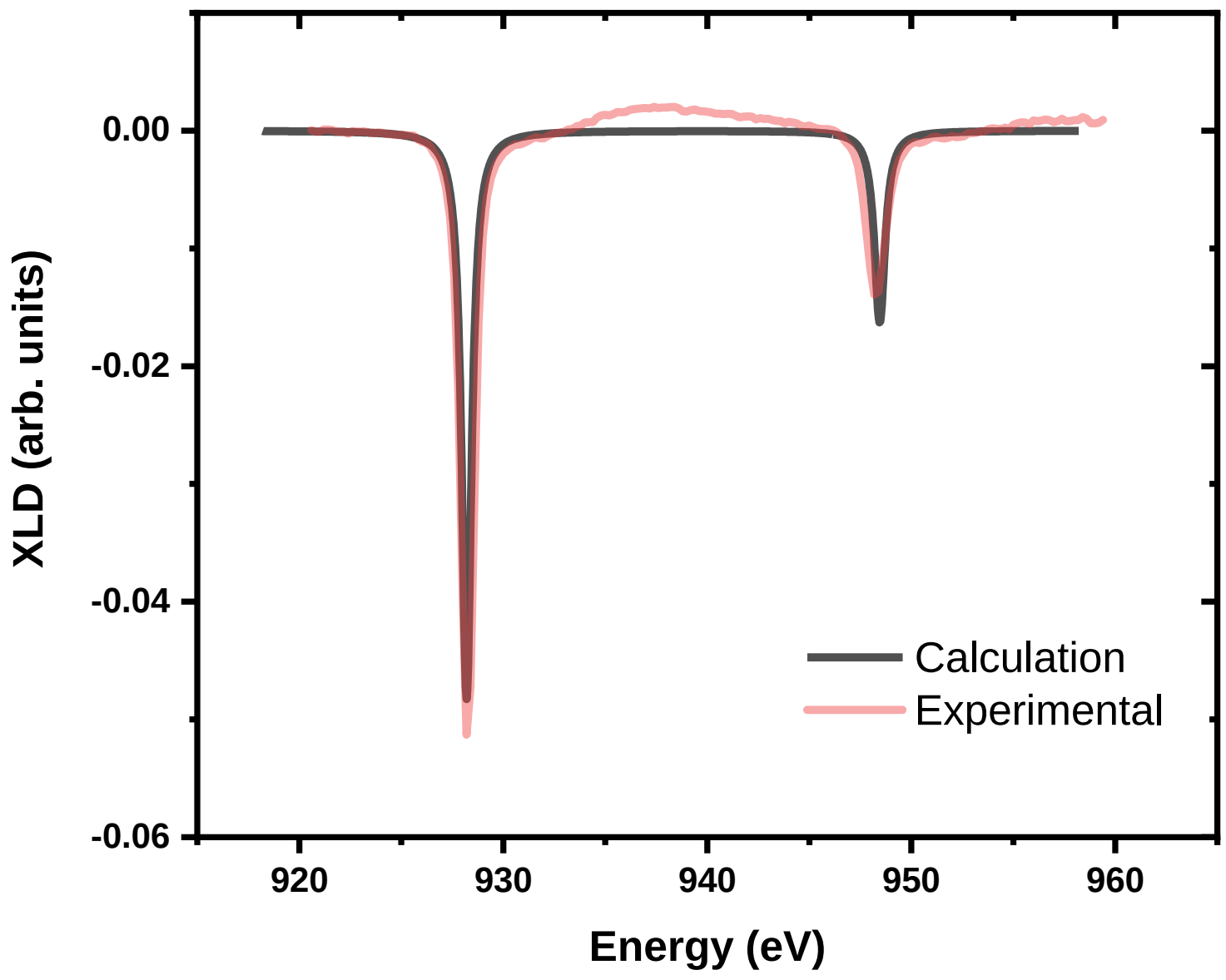

*Figure 4. Comparison of the experimental and calculated XLD signal for the C/T-CuO/STO sample. XLD signal has been collected at 300 K, with 30 deg incident angle in zero field.*

We repeat the XMCD measurements using opposite field cooling in -6.5 T from 360 K to 3 K, which results in the spectra of Figure 5. The XMCD signal maintains the same orientation to those found in the previous case of +FC (Figure 3). However, the XMCD signal for Figure 5 (b) with 6.5 T applied field remains larger than the (d) case with -6.5 T applied field, but the difference between (b) and (d) is smaller in size compared to the +FC case (Figure 3 (b-d)). This indicates that some pinned moments still point along the +H direction, i.e. cannot be reversed even at high temperatures. The corresponding sum rule results for the -FC case are also shown in Table I.

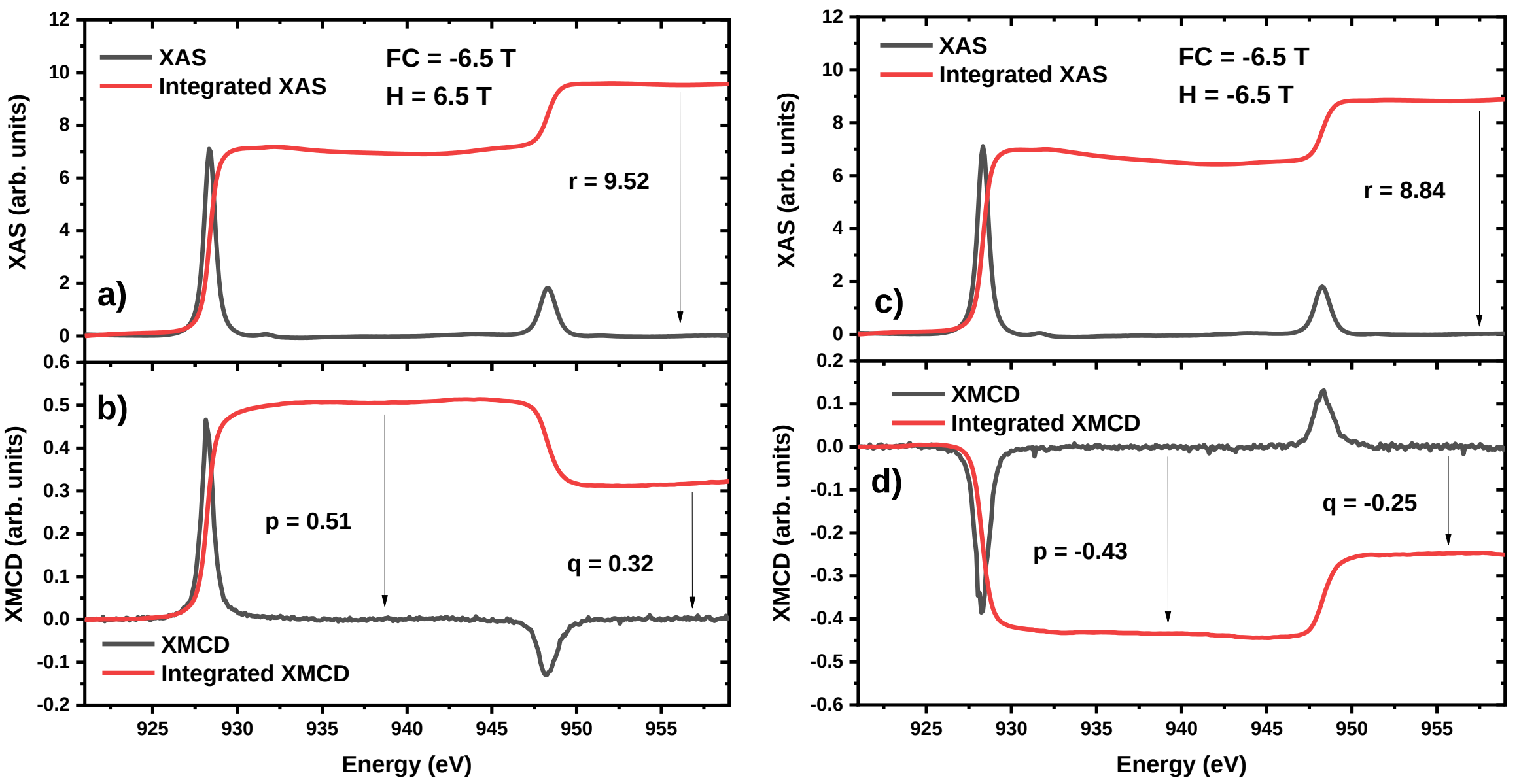

*Figure 5. XAS and XMCD signals at 3K for negative FC (-6.5 T) and applied magnetic field of 6.5 T in (a-b) and -6.5 T in (c-d), respectively. Data collected at 30 degrees incident angle.*

| | **H = 6.5 T**<br>**[$\mu_B$/atom]** | **H = -6.5 T**<br>**[$\mu_B$/atom]** |
|---|---|---|
| **+FC (6.5 T)** | $m_{orb} = -0.032 \pm 0.003$<br>$m_{spin} = -0.10 \pm 0.01$<br>$M_{total} = -0.13 \pm 0.01$ | $m_{orb} = 0.011 \pm 0.001$<br>$m_{spin} = 0.10 \pm 0.01$<br>$M_{total} = 0.11 \pm 0.01$ |
| **- FC (-6.5 T)** | $m_{orb} = -0.022 \pm 0.002$<br>$m_{spin} = -0.093 \pm 0.009$<br>$M_{total} = -0.12 \pm 0.01$ | $m_{orb} = 0.019 \pm 0.002$<br>$m_{spin} = 0.091 \pm 0.009$<br>$M_{total} = 0.11 \pm 0.01$ |

*Table I. Orbital, spin and total magnetic moment of $Cu^{2+}$ in T-CuO extracted from XMCD sum rules around the Cu $L_{2,3}$ edges at 3K. FC from 360 K to 3 K in a field of either 6.5 T or -6.5 T and data taken in applied field of 6.5 T or -6.5 T.*

The absolute value for the spin moment obtained through XMCD sum rules is approximately 0.1 $\mu_B$/atom for both opposite FC cases. However, the absolute value of the orbital magnetic moment for the +FC case differ by more than a factor two between 6.5 T and -6.5 T. This is not the case for -FC, where the spin and orbital moments are equal, within the uncertainty, for both applied fields. On the other hand, we find that the total magnetic moment is equal in all cases with approximately 0.12 $\mu_B$/atom. Moreover, we note that the spin and orbital moments conserve same sign for the same field applied independently of the FC direction.

Figure 6 shows the XMCD signal taken at the maximum of the Cu $L_3$ edge as a function of applied magnetic field following the two opposite FC processes. The XMCD signal is independent of the field ramping direction and does not show any hysteresis, indicative of an absence of an intrinsic ferromagnetic state. The XMCD signal exhibits a clear positive shift in y-axis from the origin. The shift indicates that a significant fraction of the magnetic moments does not reverse their orientation with the applied magnetic field at low temperature, a clear indication that these moments remain pinned. The estimated amount of pinned moments is approximately 37% of the total moment ($M_{\mathrm{pin}}$+$M_{\mathrm{rot}}$) observed in the +FC case (Figure 6a). The measured data are compared with the calculated magnetization (dashed blue curve) based on a paramagnetic system with spin ½ in an applied field at 3 K without any pinned moment contribution. The calculation represents simply the Brillouin function [28]. The excellent agreement of our data removing the pinned contribution with this simple model indicates a mixture of paramagnetic and pinned magnetic Cu moments.

Figure 6b shows the results for the -FC case. Interestingly, again a positive shift in y-axis is observed, although significantly smaller, reflecting the smaller contribution from pinned moments that account for 16% of the total observed moment. This indicates that not all moments that are pinned at low temperature reverse at 360 K in -6.5 T field and those pinned moments that do not reverse have a preferred orientation.

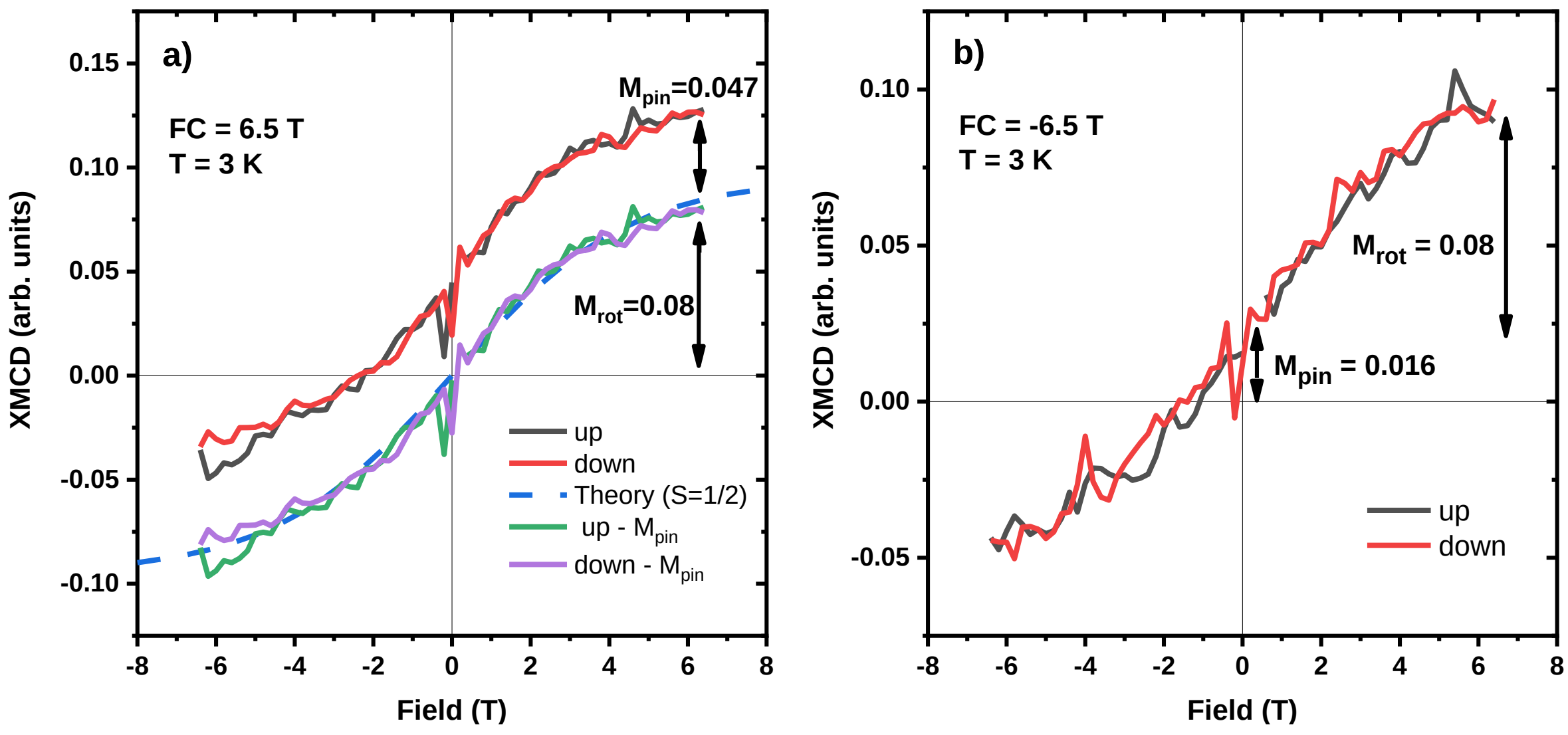

*Figure 6. XMCD signal at the maxima of the Cu $L_3$ edge as a function of ramped magnetic field measured at 3 K following (a) +FC in +6.5 T and (b) -FC in -6.5 T. $M_{pin}$ and $M_{rot}$ indicate the contribution of pinned and rotated moments, respectively. The dashed line in (a) represents the Brillouin function fo r spin ½. Data collected at a 30 degrees incident angle.*

In addition, Figure 7 shows the temperature dependence of the XMCD signal at the maximum of Cu $L_3$ edge. We find a reasonable agreement between the partition function $Z$ of a paramagnetic system with spin ½ [28].

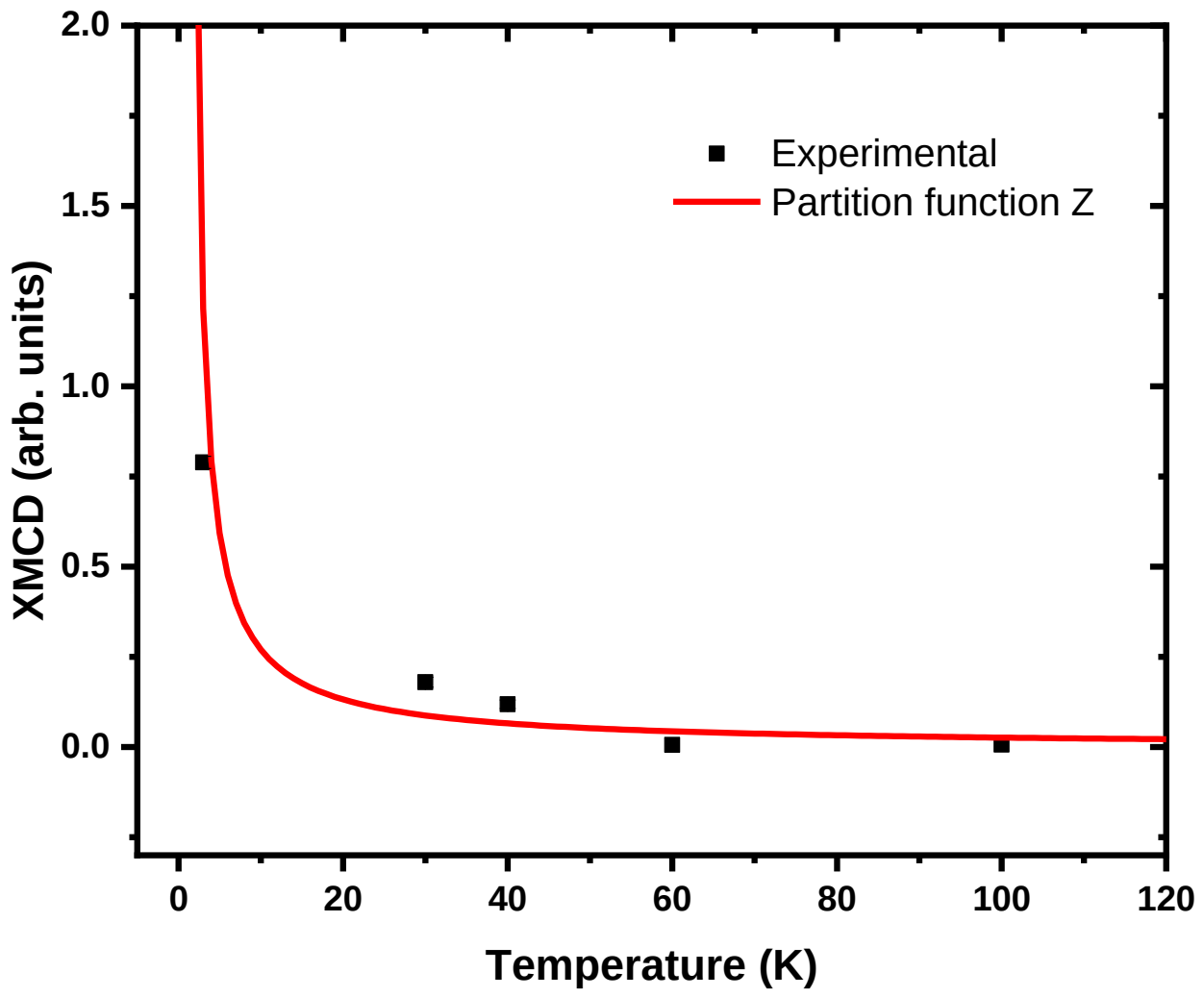

*Figure 7. Experimental data of the XMCD signal at the maximum of Cu $L_3$ edge as a function of temperature (dark squares) and the partition function Z of a paramagnetic system with spin ½ (red line). Data collected at a 30 degrees incident angle.*

In order to study the direction of the pinned moments, Figure 8 displays an XMCD hysteresis loop performed at normal incidence angle. The sample was FC in -6.5 T from 360 K to 3 K. Contrary to the previous case, Figure 8 does not exhibit a vertical shift, indicating that the pinned moments do not have any out-of-plane contribution, and therefore, the XMCD signal consists of only the paramagnetic contribution. This confirms that the pinned moments are parallel to the sample surface. Moreover, the paramagnetic signal is of comparable size to that probed at 30 degrees grazing angle (Figure 6a), which shows that the paramagnetic spins are isotropic.

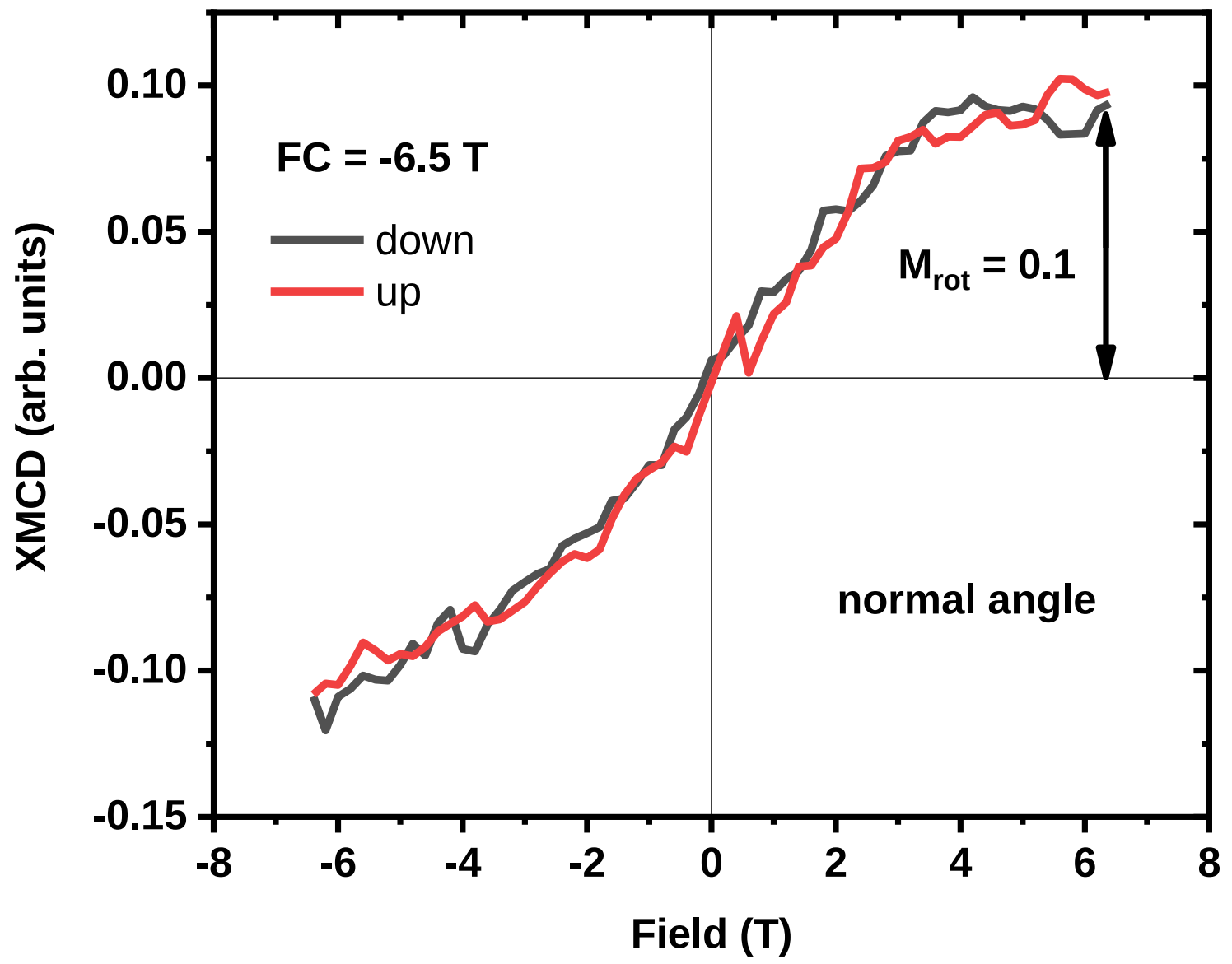

*Figure 8. XMCD as a function of applied field at normal incidence angle. FC of -6.5 T. The arrow indicates the contribution of rotated moments ($M_{rot}$).*

Furthermore, to identify the contribution of pinned moments to the signal, XMCD measurements in zero-field are shown in Figure 9. These measurements indicate that the pinned moments have a larger orbital contribution compared to the spin contribution, a similar case was reported in [29]. Applying the sum rules, we find that the pinned moments exhibit a ratio of $m_{orb}/m_{spin} \approx 2$ and a total magnetic moment of 0.017±0.001 $\mu_b$/atom. Comparing these values to the total magnetic signal observed at applied field measurements, $M_{total}$ *0.12* $\mu_b$/atom (Table I), we deduce that the pinned moments represent approximately 15%, in agreement with our above estimation from Figure 6b.

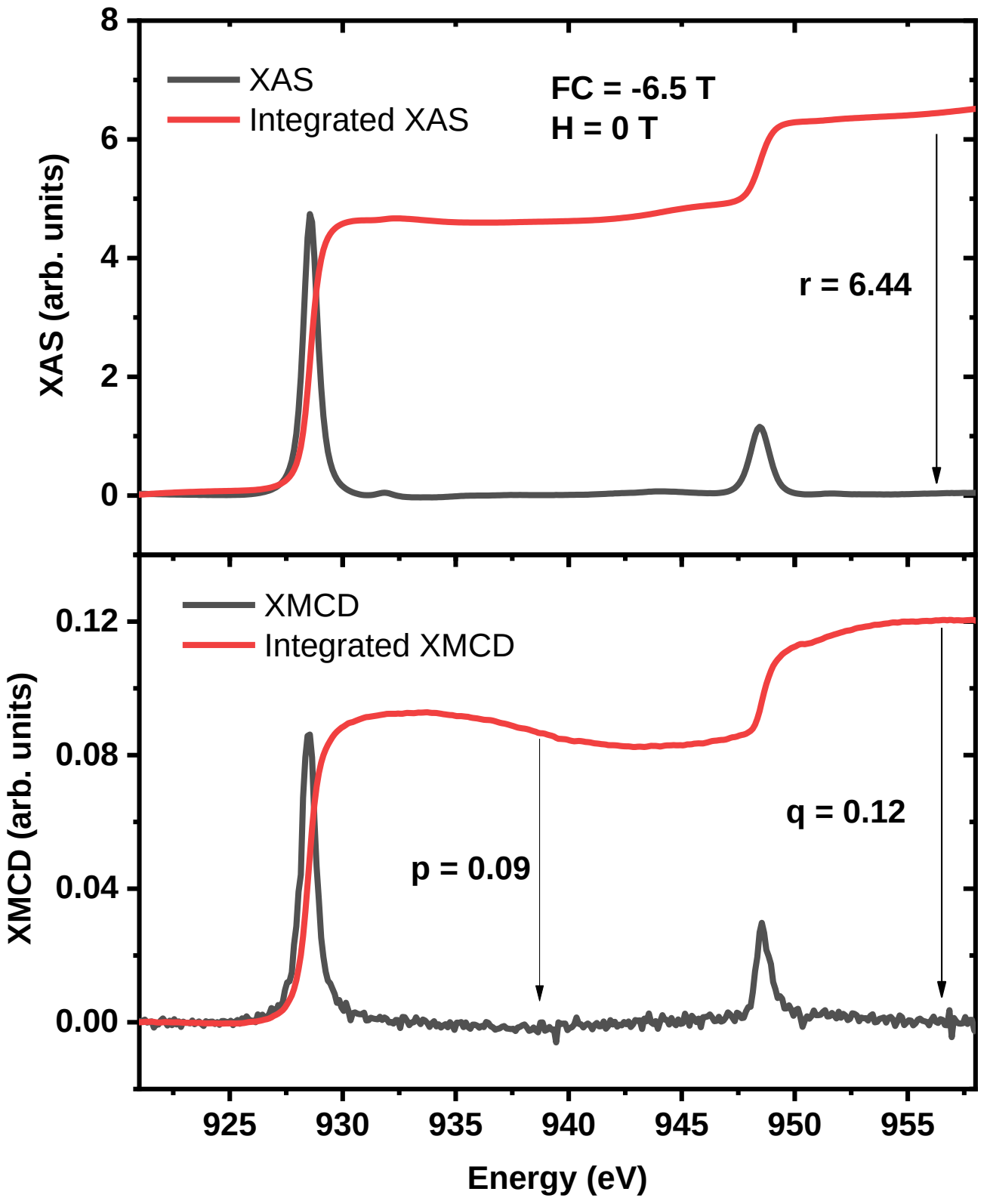

*Figure 9. (Top) XAS and (Bottom) XMCD data measured at Cu $L_{2,3}$ edges. FC in -6.5 T from 360 K to 3 K and data taken in zero field at 3K. Data taken at an incident angle of 30 degrees.*

As the paramagnetic and the pinned moment are most likely located at the interfaces, it would be interesting to check if there is a magnetic moment at the Ti ions from the STO substrate. Such moments have been already reported in the presence of a two-dimensional electron gas (2DEG) [30]-[33]. The XAS and the XMCD signals obtained at the Ti $L_{2,3}$ edges are shown in Figure 10. Only a very small XMCD signal is observed around the Ti $L_{2,3}$ edges, which is similar to those found in annealed LAO/STO samples [30]. However, we do not observe any response from magnetic $Ti^{3+}$ ions similar as found for the non-annealed LAO/STO [30] or in γ-$Al_2O_3$/$SrTiO_3$ [33], both exhibiting a 2DEG. This indicates that the T-CuO/STO interface is unlikely to host a 2DEG gas and that the observed XMCD is not due to an accumulation of $Ti^{3+}$ or any associated magnetic order at the STO interface as reported in other systems [30]-[33].

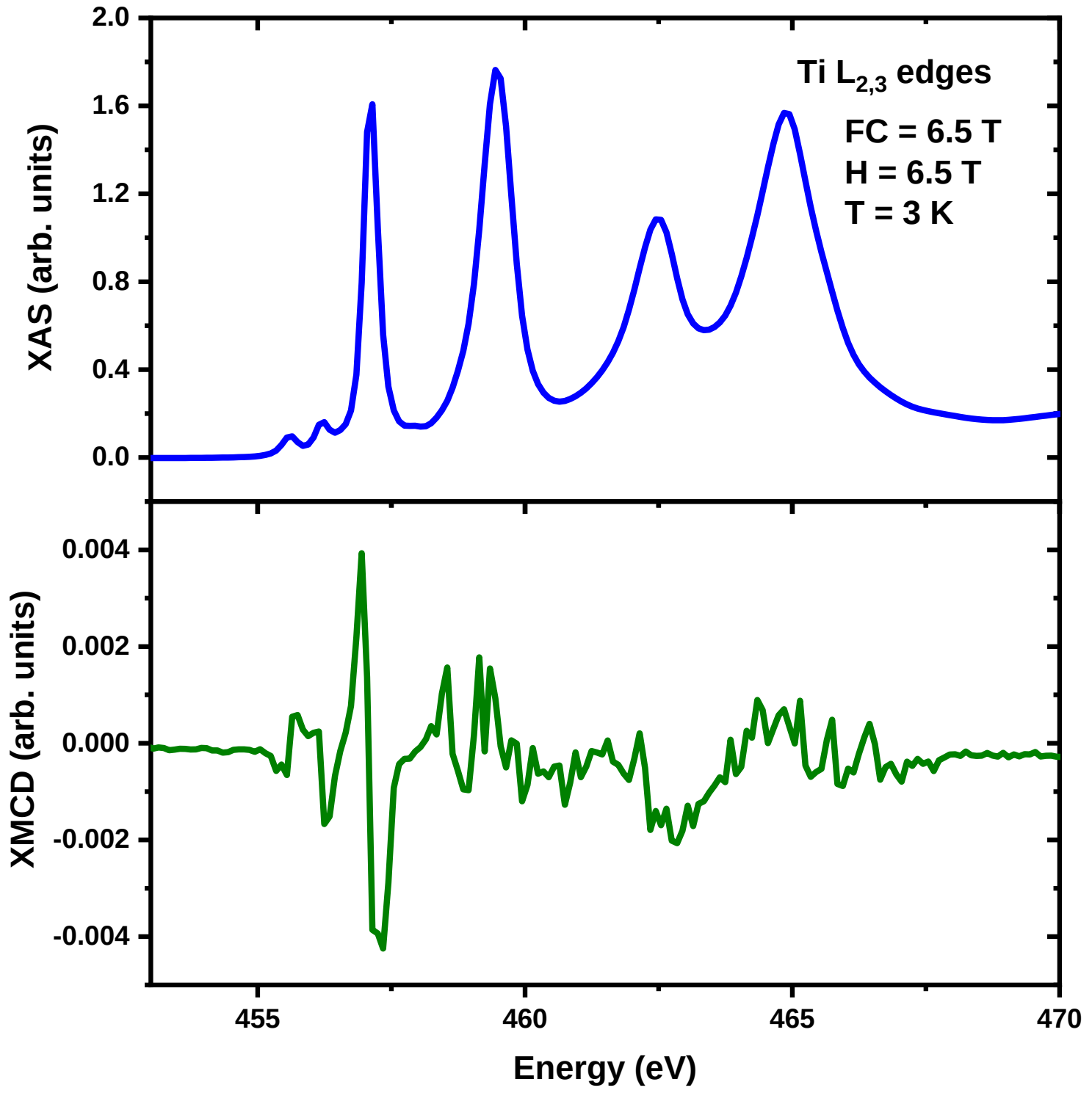

*Figure 10. XAS (top) and XMCD (bottom) signal taken at the Ti $L_{2,3}$ edge with +FC, in an applied magnetic field of 6.5 T and T=3 K.*

## 4 Discussion & conclusions

The relatively large fraction of depolarized muons implanted in the C/T-CuO/STO sample compared to the C/STO sample throughout the investigated temperature range gives a clear indication that a significant portion of the C/T-CuO/STO film is strongly magnetic. The 6% reduction of asymmetry in comparison with C/STO corresponds to at least 1/3 of the volume fraction contribution of the muons stopping in the T-CuO layer, predicted from our very rough Trim.SP calculations (~18%). This indicates that at least 1/3 of the T-CuO volume has a long-range magnetic order. This is also supported by our depth dependent measurements in Figure 2b. It is also important to point out here the fact that the lost asymmetry does not exceed what we expect from the T-CuO layer, evidencing that the magnetic order in this layer cannot be ferromagnetic. Strong static stray fields from a thin ferromagnetic layer would have caused a much larger loss of asymmetry, larger than the volume fraction of the layer itself [35]. In addition, the absence of noticeable magnetic field shift in Figure 2d confirms that T-CuO cannot be ferromagnetic, since this would lead to a negative shift relative to the non-magnetic C/STO [35].

Finally, the enhancement of $\sigma$ is another indication of the magnetic field fluctuations experienced by muons stopping in the vicinity of the T-CuO layer. These observations together support a scenario where the T-CuO layer orders antiferromagnetically (or a magnetic order with very small net magnetic moment) and constitute a first direct observation of the magnetic order in T-CuO.

The field dependent XMCD data of T-CuO exhibit two magnetic contributions, one coming from isotropic paramagnetic $Cu^{2+}$ moments and another from pinned moments lying in the sample surface plane. We have observed that the pinned moment contribution strongly depends on the field direction in FC procedure starting at 360 K. This indicates that the pinned moment contribution originates in two portions/fractions: "permanently" and "eventually" pinned moments. The "permanently" pinned moments remains even at 360 K and the "eventually" pinned moments are reversible at 360 K but are frozen at 3 K. The XMCD sum rules give an estimation of the total magnetic moment of around 0.12 $\mu_B$/atom, from which approximately 15% of the total magnetic moment corresponds to permanently pinned moments. The contribution of pinned moments can be enhanced up to 37% by the "eventually" pinned moments depending on the FC direction. The observed $Cu^{2+}$ magnetic moment of 0.12 $\mu_B$/atom is only 1/5 of the total magnetic moment obtained by the multiplet calculations ($S_Z+L_Z$= 0.615 $\mu_B$/atom). Taking into account that the T-CuO layer consists of only 6 u.c., the magnetic signal observed comes from just approximately 1.2 u.c. of T-CuO.

The simplest model to explain our observations is schematically presented in Figure 11. It is common that an AFM system shows uncompensated moments at the surface, e.g., due to uncompensated u.c., surface roughness and/or defects and domain walls. All of them lead to missing magnetic exchange paths from the discontinuity of the magnetic ordering. We have shown that T-CuO exhibits uncompensated moments and with its major fraction being paramagnetic (PM) even at low temperature. Our XMCD data indicate that the pinned moments have enhanced orbital moment contribution which one could expect from inversion symmetry breaking or reduced dimensionality, e.g., as found for increased angular magnetic moments at a surface [34]. In addition, we have demonstrated that the pinned moments are oriented along the sample surface. One would assume that the orientation of the pinned moments reveals the direction of the AFM ordering. That would imply an AFM ordering of ferromagnetic planes along (001). The FC process enhances the pinned moment layer when it is along the pinned moment direction. The fact that a significant fraction of the pinned moments has same preferred orientation (i.e., "permanently" pinned moments), independently of the direction of FC applied after warming up to 360 K, is a clear indication that the AFM ordering is present even at 360 K, which represents a lower limit for the transition temperature in T-CuO.

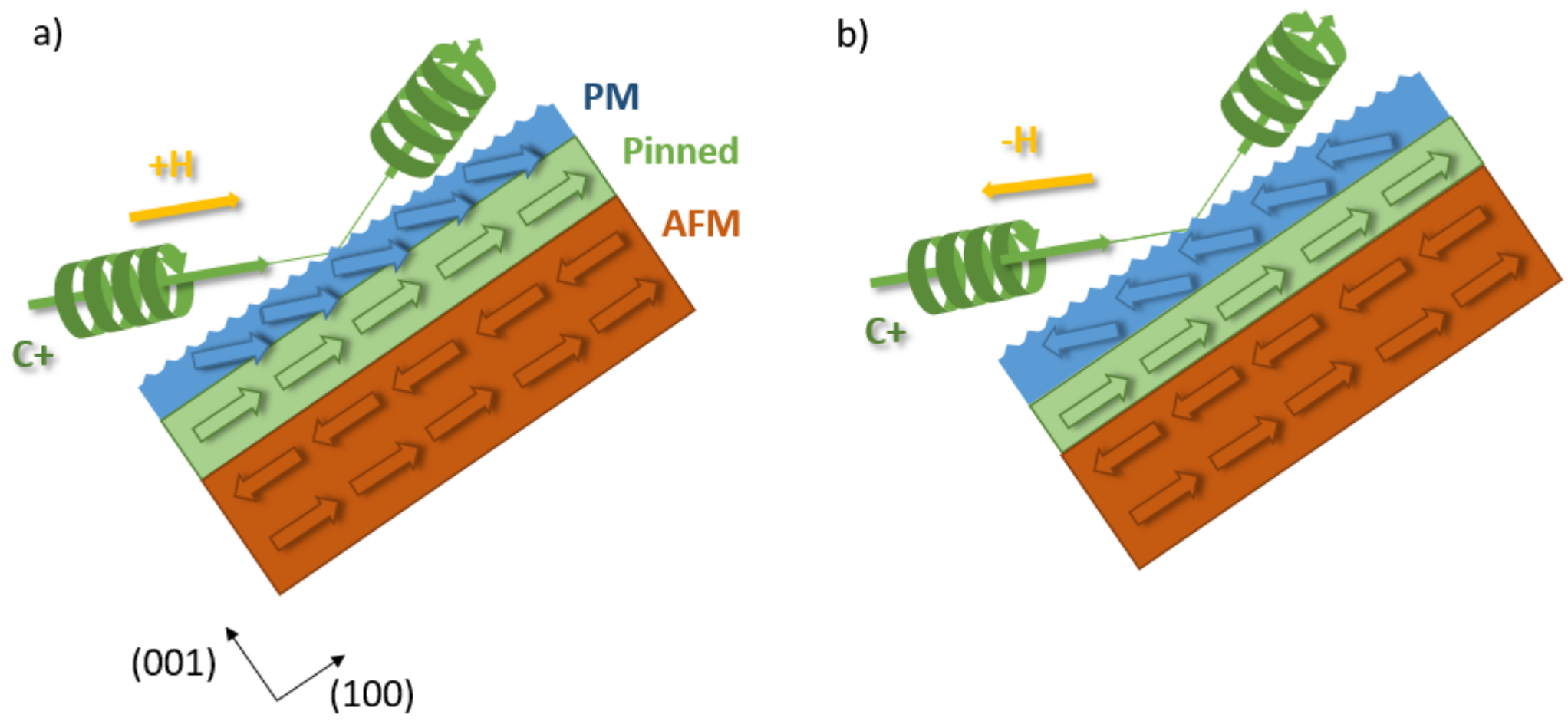

*Figure 11. Sketch of the proposed model for T-CuO. (a) represents the + FC case and (b) represents the -FC case. The blue layer highlights the paramagnetic moments. The green layer corresponds to the pinned moments and the orange one to the antiferromagnetic moments. Note that in (a) the pinned moment layer is enhanced because the pinned moment and paramagnetic moments (PM) have same orientation.*

Although our simplistic model explains all the phenomena observed in our study, an AFM ordering of FM planes along the (001) direction does not agree with the theoretical prediction in ref. [8]. The magnetic exchange interaction between the $Cu^{2+}$ ions avoids having ferromagnetic single layers as the orange layer sketched in our model (Figure 11). Nevertheless, T-CuO is highly frustrated making it more difficult to predict theoretically the nature of magnetic ordering. In addition, missing magnetic exchange interactions from defects at the surface makes the assumption that the pinned moments are parallel to the AFM moments, questionable. However, the fraction of pinned moments (16 % of the total magnetic moment observed) is significant and it might be difficult to solely attribute them to defects.

Summarizing, our LE-μSR study supports an AFM ordering in T-CuO with a transition temperature higher than 200 K. The XMCD investigation exhibits a magnetic signal constituted by pinned moments oriented along the sample surface and isotropic paramagnetic moments. The origin of the isotropic paramagnetic moments is most likely roughness, surface defects and/or uncompensated spins. In addition, the detection of pinned moments lying in the film plane and remaining up to 360 K, gives clear indication of a bias that originates from an underlying AFM order.

**Acknowledgments:**

Part of this work was performed at the Swiss Muon Source SμS, Paul Scherrer Institute, Villigen, Switzerland. XMCD measurements were performed on the EPFL/PSI X-Treme beamline at the Swiss Light Source, Paul Scherrer Institute. We want to acknowledge D. K. for performing the carbon coating. We also thank to Prof. Frederic Mila for helpful discussions on the theoretical understanding of the T-CuO system. We thank the X11MA and μE4 beamline staff for experimental support. N. O. H. acknowledges financial support of the Swiss National Science Foundation, No. 200021_169017. J. R. L. M. was supported by the National Centers of Competence in Research in Molecular Ultrafast Science and Technology (NCCR MUST No. 51NF40-183615) from the Swiss National Sciences Foundations. S.M. acknowledges support by the Swiss National Science Foundation (Grant No. P300P2-171221). This research used resources of the Advanced Light Source, which is a DOE Office of Science User Facility under Contract No. DE-AC02-05CH11231.